\documentclass[preprint,showpacs,preprintnumbers,superscriptaddress]{revtex4-1}
\usepackage{multirow}
\usepackage{diagbox}
\usepackage{booktabs}
\usepackage{array}
\usepackage{natbib}
\usepackage{amsmath,amssymb,graphicx,bm,subfigure,float,siunitx,color}
\usepackage{bm,subfigure,float,siunitx,graphicx}
\usepackage[figuresright]{rotating}
\usepackage{color}
\usepackage{epstopdf}
\usepackage{mathrsfs}
\newcommand{\be}{\begin{equation}}
\newcommand{\ee}{\end{equation}}

\newcommand{\1}{\left}
\newcommand{\2}{\right}
\def\({\left(}
\def\){\right)}
\def\[{\left[}
\def\]{\right]}

\newcommand{\dif}{\,\mathrm{d}}

\newcommand{\me}{\mathrm{e}}

\newcommand{\p}{\partial}

\newcommand{\m}{\mu}
\newcommand{\n}{\nu}

\renewcommand{\th}{\theta}

\newcommand{\na}{\nabla}

\begin{document}
\title{\boldmath  The perturbation solutions to the Blandford-Znajek mechanism in the Kerr-Sen black hole}
\author{Haiyuan Feng\footnote{Corresponding author}}
\email{Email address:  fenghaiyuan@sxnu.edu.cn}
\affiliation{School of Physics and Electronic Engineering, Shanxi Normal University, Taiyuan 030031, China}

\author{Ziqiang Cai}
\email{Email address: gs.zqcai24@gzu.edu.cn}
\affiliation{College of Physics, Guizhou University, Guiyang 550025, China}

\author{Rong-Jia Yang}
\email{Email address: yangrongjia@tsinghua.org.cn}
\affiliation{College of Physical Science and Technology, Hebei University, Baoding 071002, China}

\author{Jinjun Zhang\footnote{Corresponding author}}
\email{Email address: zhangjinjun@sxnu.edu.cn }
\affiliation{School of Physics and Electronic Engineering, Shanxi Normal University, Taiyuan 030031, China}

\begin{abstract}


We investigate the steady, axisymmetric, force-free magnetosphere of Kerr–Sen black hole (BH) within the framework of the Einstein–Maxwell–dilaton–axion (EMDA) theory. By perturbatively solving the nonlinear Grad–Shafranov (GS) equation, we determine the magnetic field configuration and quantify the influence of the dilaton parameter $r_2$ on the energy extraction rate and radiative efficiency. Our results show that both the energy extraction power and the radiative efficiency increase with $r_2$, exceeding those of the standard Kerr BH, whereas the extraction efficiency remain consistent with the Kerr case. In addition, we perform $\chi^2$ statistical analysis using observational data from six binary BH systems, which indicates that the Kerr BH currently provides a better fit for bulk Lorentz factors $\Gamma = 2$ and $5$.
\end{abstract}

\maketitle
\section{Introduction}
Accretion disks surrounding black holes (BHs) are ubiquitous in the universe. These systems often launch relativistic jets that carry a substantial portion of the accretion energy. In BH binaries, two principal types of jets have been identified \cite{Fender:2004gg}: steady jets, which appear across a wide range of luminosities in the hard state, and transient jets, which emerge during transitions between spectral states. Evidence also suggests that transient jets can occur during the return from the soft to the hard state at low accretion rates, preceding a quiescent phase \cite{Belloni:2016xgi}. 
Despite extensive observations, the physical mechanism driving jet formation remains uncertain. Within the framework of force-free electrodynamics, Blandford and Znajek (1977) proposed that the rotational energy of spinning BH can be extracted via electromagnetic fields penetrating the event horizon, producing Poynting flux dominated outflow \cite{10.1093/mnras/179.3.433,Lee:1999se}.  Hence, the Blandford-Znajek (BZ) mechanism provides a robust theoretical explanation for the phenomenon of jets in astronomy.

Analytical and numerical investigations have extensively examined the structure of magnetic fields around BHs. A self-consistent description of the highly magnetized plasma in rotating BHs requires solving the nonlinear Grad-Shafranov (GS) equation. The GS equation serves as the master equation governing the equilibrium structure of a stationary, axisymmetric, and force-free magnetosphere surrounding BH. It determines the distribution of the $A_{\phi}$ by coupling the magnetic field geometry with the system's conserved quantities. In curved spacetime, the GS equation possesses three regular singular surfaces: the event horizon and the inner and outer light surfaces, the latter marking the locations where the rotational velocity of magnetic field lines approaches the speed of light. To maintain the global smoothness of the magnetic field in non-extremal configurations, appropriate boundary conditions must be imposed at each of surfaces. Nevertheless, the mathematically rigorous proof for the existence of $C^1$ solutions remains an open challenge. This provides a strong motivation for adopting the perturbation method \cite{1,2,3,4,5}. BZ first employed perturbation techniques to analytically solve the GS equation and obtained a solution for split monopole magnetic field. However, due to the intricate nature of the nonlinear GS equation, perturbation techniques have not seen further advancement to date. Futhermore, simulations in general relativistic magnetohydrodynamics (GRMHD) and magnetodynamics (GRMD) provide us with an opportunity to explore the BZ mechanism. GRMHD simulations \cite{McKinney:2004ka, McKinney:2005zw} indicate that, in the polar region, the monopole perturbation solution provides a good description of the magnetic field configuration as well as the angular distribution of energy flow. Additionally, GRMD simulations illustrated in Ref. \cite{article,Komissarov:2004ms} demonstrate that the analytical monopole solution aligns well with numerical simulations, especially for slowly rotating BHs. Since then, the BZ mechanism has been employed as one of the most crucial models in astrophysics to elucidate the phenomenon of jets.


While General Relativity (GR) remains the standard framework for BHs, jet phenomena are increasingly investigated within alternative gravity to address GR's inherent limitations \cite{Pei:2016kka}. Theoretically, the existence of singularities and the resulting loss of predictability \cite{Penrose:1964wq, Hawking:1976ra, Christodoulou:1991yfa} suggest that GR is an incomplete description, necessitating modifications that incorporate quantum gravity effects \cite{Will:2014kxa}. Observationally, GR faces significant challenges in accounting for dark matter and dark energy phenomena essential for explaining galactic rotation curves and the universe's accelerated expansion \cite{Milgrom:1983pn, Bekenstein:1984tv, Milgrom:2003ui, SupernovaSearchTeam:1998fmf, Clifton:2011jh}.

In the realm of alternative theories, the Einstein-Maxwell-dilaton-axion (EMDA) model has attracted considerable attention \cite{Rogatko:2002qe, Sen:1992ua}. This model integrates the dilation field and the pseudoscalar axion, intricately linked to the metric and the Maxwell field. The origins of the dilaton and axion fields trace back to string compactifications, resulting in significant implications for inflationary cosmology and the late time acceleration of the universe \cite{Catena:2007jf, Sonner:2006yn}. Futhermore, exploring the role of such a theory in astrophysical observations holds substantial value. Within these string-inspired low-energy effective theories, understanding the constrained parameter range becomes pivotal. For example, a preferred value of $r_2\equiv\frac{Q^2}{M}\thickapprox0.2M$ is determined based on the optical continuum spectrum of quasars \cite{Banerjee:2020qmi}. Additionally, recent investigations have established observational constraints on the dilaton parameter ($0.1M\lesssim r_2\lesssim0.4M$) by analyzing the shadow diameters of M87* and Sgr A* \cite{Sahoo:2023czj}. These result highlight the credibility of the EMDA model in the field of astrophysics. Moreover, the model could reveal the feasibility of high-energy string theory in making physical predictions.

The article is organized as follows: In Section II, we will briefly review the EMDA model and the Kerr-Sen BH. In Section III, based on the assumption of force-free and convergence condition, we will derive second-order perturbation solutions for the GS equation. Subsequently, we will analyze various physically observable quantities, including BH energy extraction rate, extraction efficiency, BH radiative efficiency, and ergosphere, within the framework of the Novikov-Thorne model. Furthermore, we will compare the differences between Kerr and Kerr-Sen BHs. In Section IV, we will examine a dataset consisting of six binary BHs, with a specific focus on radiative efficiency and energy extraction rate. Additionally, we will use chi-square distribution to calculate the optimal parameter $r_2$ for Kerr-Sen BH. Summary and discussion are given in Section V. For convenience, we will use geometrical units $c=G=1$ and signature convention $(-,+,+,+)$ for spacetime metric throughout the article.

\section{Kerr-Sen black hole in Einstein-Maxwell-Dilaton-Axion gravity}
The EMDA model arises as the low-energy effective description of heterotic string theory, offering a compelling theoretical framework to explore the intricate interactions among gravitational, electromagnetic, and scalar fields. Within this theory, the primary dynamical fields include the spacetime metric $g_{\m\n}$, which encodes the geometry of the background manifold; the gauge vector field $a_{\m}$ describing the electromagnetic interaction; the dilaton field $\chi$, a scalar field that modulates the effective coupling between matter and gravity; and the axion field $\xi$, a pseudo-scalar field emerging from string compactification, which changes sign under parity transformation and couples to the electromagnetic dual tensor, thereby introducing parity-violating effects \cite{Rogatko:2002qe,Sen:1992ua,Campbell:1992hc}. 

The action of EMDA model can be systematically derived by combining the bosonic sector of supergravity with the gauge sector of super–Yang–Mills theory, leading to an effective action 
\be
\label{1}
S=\frac{1}{16\pi}\int\sqrt{-g}d^4x\[R-2\p_{\m}\chi\p^{\m}\chi-\frac{1}{2}\me^{4\chi}\p_{\m}\xi\p^\m\xi+
\me^{-2\chi}f_{\m\n}f^{\m\n}+\xi f_{\m\n}\widetilde{f}^{\m\n} \],
\ee
where $R$ represents the Ricci scalar, $ f_{\mu\nu}$ denotes the second-order antisymmetric Maxwell field strength tensor, defined as $f_{\mu\nu} = \nabla_{\mu} a_{\nu} - \nabla_{\nu} a_{\mu}$. The term $\widetilde{f}^{\mu\nu}$ corresponds to the dual of the field strength tensor, encapsulating the magnetic components of the electromagnetic field. These quantities together describe the coupling between the electromagnetic and gravitational sectors, as well as their interaction with the dilaton and axion fields in the EMDA framework.

By varying the action with respect to the metric, the dilaton, axion, and electromagnetic fields, one obtains the following equations of motion
\be
\1\{\begin{split}
\label{2}
&\Box\chi-\frac{1}{2}\me^{4\chi}\na_\m\xi\na^\m\xi+\frac{1}{2}\me^{-2\chi}f_{\m\n}f^{\m\n}=0,\\
&\Box\xi+4\na_\m\xi\na^\m\xi-\me^{-4\chi}f_{\m\n}\widetilde{f}^{\m\n}=0,\\
&\na_\m\widetilde{f}^{\m\n}=0,\\
&\na_\m(\me^{-2\chi}f^{\m\n}+\xi\widetilde{f}^{\m\n})=0,\\
&G_{\m\n}=\me^{2\chi}(4f_{\m\rho}f^{\rho}_{\n}-g_{\m\n}f^2)-g_{\m\n}(2\na_\m\chi\na^\m\chi+\frac{1}{2}\me^{4\chi}\na_\m\xi\na^\m\xi)\\
&+\na_\m\chi\na_\n\chi+\me^{4\chi}\na_\m{\xi}\na_\n{\xi}.
\end{split}\2.
\ee
It can be seen that the dilaton, axion, electromagnetic, and gravitational fields are inherently coupled within the EMDA framework. It should be emphasized that $\na_\m\widetilde{f}^{\m\n}=0$ merely expresses the Bianchi identity. It corresponds to the constraint equation in Maxwell equation and does not represent any independent dynamical degrees of freedom. Furthermore, the explicit solutions for the axion, dilaton, and $U(1)$ gauge field can be found in Refs. \cite{Sen:1992ua,Campbell:1992hc,Ganguly:2014pwa}, which follows 
\be
\1\{\begin{split}
\label{3}
&\xi =\frac{q^2}{\mathcal{M}} \frac{a \cos \theta}{r^2+a^2 \cos ^2 \theta}, \\
&e^{2 \chi} =\frac{r^2+a^2 \cos ^2 \theta}{r\left(r+r_2\right)+a^2 \cos ^2 \theta},\\
& A=\frac{q r}{\Sigma}\left(-\mathrm{d} t+a \sin ^2 \theta \mathrm{~d} \phi\right).
\end{split}\2.
\ee

It follows from Eq. \eqref{3} that all three fields vanish in the asymptotic limit $r \to \infty$ and, in this region, exhibit properties analogous to those of GR. Moreover, Eq. \eqref{3} also show that the coupling between the axion–dilaton sector and the Maxwell field is essential, as in its absence the associated field strengths would vanish identically. Therefore, although the BH carries electric charge, this charge originates from the axion–photon coupling rather than from infalling charged matter. Further analytical manipulations lead to the axisymmetric solution in Kerr–Schild coordinates. This solution, known as the Kerr–Sen black hole, represents \cite{Garfinkle:1990qj}
\be
\label{4}
\begin{split}
\dif s^{2}&=-\(1-\frac{2M r}{\tilde{\Sigma}}\)\dif t^{2}+\(1+\frac{2Mr}{\tilde{\Sigma}}\)\dif r^{2}+ \frac{4Mr}{\tilde{\Sigma}}\dif t\dif r + \tilde{\Sigma}\dif \th^{2} -\frac{4aMr\sin^2{\th}}{\tilde{\Sigma}}\dif t\dif\phi\\
&+\sin^2{\th}\dif \phi^2\(\tilde{\Delta}+\frac{2Mr(r(r+r_2)+a^2)}{\tilde{\Sigma}}\)-2a\(1+\frac{2Mr}{\tilde{\Sigma}}\)\sin^2{\th}\dif r\dif\phi ,
\end{split}
\ee
with
\be
\1\{\begin{split}
\label{5}
&\tilde{\Sigma}=r(r+r_2)+a^2\cos^2{\th},\\
&\tilde{\Delta}=r(r+r_2)-2Mr+a^2,
\end{split}\2.
\ee
here $M$ denotes the BH mass parameter, while the dilaton parameter is defined as $r_2 = \frac{Q^2}{M}$, which is directly related to the electric charge $ Q$. When the rotation parameter $a$ is set to zero, Eq.~\eqref{4} reduces to a static, spherically symmetric dilaton BH characterized solely by its mass, charge, and asymptotic behavior \cite{Garfinkle:1990qj}. In the limiting case where the dilaton parameter $r_2$ vanishes, the Kerr–Sen geometry continuously reduces to Kerr BH.

Additionally,the event horizons $r_{\pm}$ of the Kerr–Sen BH is determined by $\tilde{\Delta}=0$, which follows
\be
\begin{split}
\label{6}
r_{+}=M-\frac{r_2}{2}+\sqrt{\left(M-\frac{r_2}{2}\right)^{2}-a^{2}},\quad\quad
r_{-}=M-\frac{r_2}{2}-\sqrt{\left(M-\frac{r_2}{2}\right)^{2}-a^{2}}.
\end{split}
\ee
From Eq.~\eqref{6}, the condition for the presence of two distinct horizons leads to the inequality $a^{2} \leq (M - \tfrac{r_2}{2})^{2}$. This relation confines the parameters to the range $0 \leq \tfrac{r_2}{M} \leq 2(1 - \tfrac{a}{M})$. Given that the spin parameter $a$ must always remain smaller than or equal to the BH mass $M$, the resulting physically viable interval for the dilaton parameter is restricted to $0 \leq \tfrac{r_2}{M} \leq 2$.

Figure 1 shows that the shaded domain marks the range of parameters producing regular BHs ($r_{+}>r_{-}$), whereas the unshaded region corresponds to cases where the horizons either coincide or vanish, giving rise to naked singularity. 
\begin{figure}[H]
\centering
\begin{minipage}{0.5\textwidth}
\centering
\includegraphics[scale=0.7,angle=0]{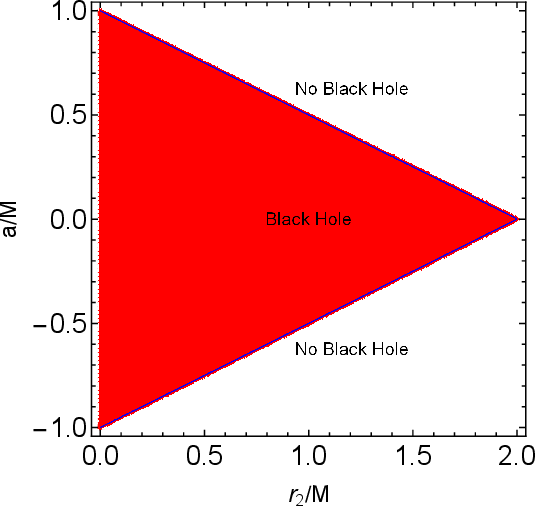}
\end{minipage}%
\caption{\label{fig.1}{The horizontal axis represents the dilaton parameter $r_2$, while the vertical axis denotes the BH spin $a$. The red region corresponds to the parameter space where two distinct event horizons exist, whereas the white region indicates the domain in which naked singularity appears.}}
\end{figure}

\section{Stationary axisymmetric force-free fields around Kerr-Sen BH}
In this section, we briefly review the fundamental equation governing stationary axisymmetric force-free fields around Kerr-Sen BH. Given the magnetic dominance of the BH's magnetosphere, we employ the force-free approximation, ensuring the electromagnetic field's dominance over matter, expressed as $T_{\mu\nu}\approx T^{(EM)}_{\mu\nu}=F_{\mu}^{\tau}F_{\nu\tau}-\frac{1}{4}g_{\mu\nu}F^{\alpha\beta}F_{\alpha\beta}$. The force-free condition ($F_{\mu\nu}J^{\mu}=0$) implies the vanishing of the electric field in the local rest-frame of the current, leading to $*F_{\mu\nu}F^{\mu\nu}=0$ \cite{10.1093/mnras/179.3.433,2014arXiv1406.4936P}. Then, one can readily demonstrate that $A_{\phi,\theta} A_{t,r} = A_{t,\theta}A_{\phi,r}$, signifying that $A_t$ is a function of $A_{\phi}$. The angular velocity of the magnetic field $\Omega(r,\theta)$ can be defined as follows \cite{2014arXiv1406.4936P}
\be
\label{7}
-\Omega\equiv\frac{\dif A_{t}}{\dif A_{\phi}}=\frac{A_{t,\theta}}{A_{\phi,\theta}}=\frac{A_{t,r}}{A_{\phi,r}}.
\ee

This definition is strictly derived from the ideal MHD or force-free condition, which requires the electromagnetic field tensor $F_{\mu\nu}$ to satisfy $F_{\mu\nu} u^\nu = 0$, where $u = \partial_t + \Omega \partial_\phi$ is the symmetry vector representing the rotation. Physically, this implies the existence of a reference frame rotating at angular velocity $\Omega$ in which the electric field vanishes, meaning the magnetic field lines are 'frozen' into and co-rotate with the plasma. Mathematically, expanding the $r$ and $\theta$ components of this condition leads directly to $\Omega = -A_{t,r}/A_{\phi,r} = -A_{t,\theta}/A_{\phi,\theta}$, an identity that ensures the  $A_t$ is a functional of the  $A_\phi$ ($A_t = A_t(A_\phi)$). Consequently, $\Omega$ is constant along any given magnetic field line, representing rigid rotation angular velocity.

For simplicity, we consider a stationary and axisymmetric model, implying that $F_{t\phi} = 0$, and the non-vanishing components of the antisymmetric Faraday tensor $F_{\mu\nu}$ are
\be
\1\{\begin{split}
\label{8}
&F_{r\phi}=-F_{\phi r}=A_{\phi,r}, F_{\theta\phi}=-F_{\phi\theta}=A_{\phi,\theta}\\
&F_{tr}=-F_{rt}=\Omega A_{\phi,r}, F_{t\theta}=-F_{\theta t}=\Omega A_{\phi,\theta} \\
&F_{r\theta}=-F_{\theta r}=\sqrt{-g} B^{\phi}.\\
\end{split}\2.
\ee

The aforementioned five non-zero components of $F_{\mu\nu}$ can be expressed in terms of three free functions: $\Omega(r, \theta)$, $A_\phi(r, \theta)$, and $B_\phi(r, \theta)$. According to the definition of the energy-momentum tensor, we can further deduce that $T^{\theta}_{t} = -\Omega T^{\theta}_{\phi}$ and $T^{r}_{t} = -\Omega T^{r}_{\phi}$. Utilizing these two relations, equations $\na_{\mu}T^{\mu}_{t}=0$ and $\na_{\mu}T^{\mu}_{\phi}=0$ for energy conservation and angular momentum conservation can be reformulated as $\Omega_{,r} A_{\phi,\theta}=\Omega_{,\theta} A_{\phi,r}$ and $\(\sqrt{-g}F^{\theta r}\)_{,r}A_{\phi,\theta}=\(\sqrt{-g}F^{\theta r}\)_{,\theta}A_{\phi,r}$. These two equations indicate that $\Omega$ and $\sqrt{-g}F^{\theta r}$ are functions of $A_{\phi}$  ($\Omega\equiv\Omega(A_{\phi})$ and $\sqrt{-g}F^{\theta r}\equiv I(A_{\phi})$, where $\Omega$ and $I$ are the angular velocity of magnetic field and poloidal electric current) \cite{Pan:2014bja}.
Substitute \eqref{8} and the relation $F^{\theta r}=\frac{I(A_{\phi})}{\sqrt{-g}}$ into the equation $F^{\theta r} =g^{\theta \mu}g^{r \nu}F_{\mu\nu} $, we derive
\be
\label{9}
B^{\phi}=-\frac{I\tilde{\Sigma}+\(2M \Omega r-a\)\sin{\theta}A_{\phi,\theta}}{\tilde{\Delta}\tilde{\Sigma}\sin^2{\theta}},
\ee
The relation links the toroidal magnetic field $B^{\phi}$ with functions $A_{\phi}(r,\theta)$, $\Omega(A_{\phi})$, and $I(A_{\phi})$. The remaining conservation equations in the $r$ and $\theta$ directions, denoted as $\na_{\mu}T^{\mu}_{r}=0$ and $\na_{\mu}T^{\mu}_{\theta}=0$, which actually can be expressed as
\be
\label{10}
-\Omega\[ \(\sqrt{-g}F^{t r}\)_{,r}+\(\sqrt{-g}F^{t\theta}\)_{,\theta} \]+\[ \( \sqrt{-g}F^{\phi r}\)_{,r}+\(\sqrt{-g}F^{\phi\theta}\)_{,\theta}\]+F_{r\theta}\frac{\dif I(A_{\phi})}{\dif A_{\phi}}=0,
\ee
where three functions $A_{\phi}(r,\theta)$, $\Omega(A_{\phi})$, and $I(A_{\phi})$ are interrelated by the nonlinear equation, which is also widely recognized as the GS equation \cite{Contopoulos:2012py,Uzdensky_2005}.
To further elucidate this equation, we examine two specific cases.

(a): In the Schwarzschild case with $I = \Omega = a = 0$, the GS equation is simplified to
\be
\label{11}
LA_{\phi}\equiv\(\frac{1}{\sin{\theta}}\frac{\p}{\p r}\(1-\frac{2M}{r}\)\frac{\p}{\p r}+\frac{1}{r^2}\frac{\p}{\p \theta}\(\frac{1}{\sin{\theta}}\)\frac{\p}{\p \theta}\)A_{\phi}=0.
\ee

Its Green's function $G(r,\theta;r_0,\theta_0)$ can be found by $LG(r,\theta;r_0,\theta_0)=\delta(r-r_0)\delta(\theta-\theta_0)$ \cite{Petterson:1974bt}. It is crucial to emphasize that the solution satisfies the specified boundary condition, with $G(r,\theta;r_0,\theta_0)$ being finite at $r = 2M$ and approaching zero at infinity.

(b): In the Kerr BH, there is no exact analytical solution for the GS equation. However, the equation can be approached using perturbation techniques ($A_{\phi} = A_0 + \frac{a^2}{M^2}A_2$) \cite{znajek1977black}. Subsequently, the GS equation takes the following perturbed form
\be
\1\{\begin{split}
\label{12}
&LA_{0}=0\\
&LA_{2}=-2\sin{\theta}\cos{\theta}\frac{1}{r^2}\(\frac{M}{2r}+\frac{M^2}{r^2}\),\\
\end{split}\2.
\ee

The zero and second solution are provided in Ref. \cite{Pan:2015haa}. Notably, for the zeroth-order equation, there are two distinct types of solutions: the monopole solution $A_0=-\cos{\th}$ and the collimated uniform solution $A_0=\frac{r^2}{M^2}\sin^2{\th}$. The magnetic field lines described by these two solutions are radial distribution and collimated cylindrical shape. The various form of solutions will be employed for different physical scenarios.

\subsection{second-order perturbation solution in Kerr-Sen BH}
The nonlinearity of the GS equation makes finding exact analytical solutions exceedingly difficult. To address this, Blandford and Znajek \cite{znajek1977black} initially developed a monopole perturbation method to the order of $O(a^2)$. This approach is valid when the deformation of the poloidal field, induced by 'spinning up' a non-rotating configuration, remains small. By imposing the horizon regularity condition \cite{Pan:2015haa} and the convergence constraint at infinity \cite{znajek1977black, Pan:2014bja}, we solve the perturbed GS equation specifically for the Kerr-Sen BH.

The perturbation expression can be represented as 
\be
\1\{\begin{split}
\label{13}
&A_{\phi}=A_{0\phi}+\frac{a^2}{M^2}A_{2\phi}+o(a^4)\\
&\Omega =\frac{a}{M^2}\omega_{1}+o(a^3)\\
&B^{\phi}=\frac{a}{M^4}B^{\phi}_{1}+o(a^3)\\
&\sqrt{-g}F^{\th r}=I=\frac{a}{M^2}i_{1}+o(a^3),
\end{split}\2.
\ee
where $A_{0\phi}$ and $A_{2\phi}$ represent the zero-order and second-order terms of the function $A_{\phi}$, while $\omega_1$ and $B_1^\phi$ correspond to the first-order corrections. The function $I = \sqrt{-g}F^{\theta r}$, defined in the previous subsection as the poloidal electric current, physically represents the total net current flowing through the area enclosed within the $(r, \theta)$ plane of the magnetosphere. To ensure the physical self-consistency of the force-free solution, we will impose strict smoothness and regularity requirements: all these physical quantities must remain finite and non-divergent at both the event horizon and at spatial infinity.

The zeroth-order solution can be easily derived in the Kerr-Sen BH, where $a = 0$. Namely, the corresponding equation is
\be
\label{14}
\tilde{L}A_{0\phi}\equiv\(\frac{1}{\sin{\theta}}\frac{\p}{\p r}\(1-\frac{2M}{r+r_2}\)\frac{\p}{\p r}+\frac{1}{r(r+r_2)}\frac{\p}{\p \theta}\(\frac{1}{\sin{\theta}}\)\frac{\p}{\p \theta}\)A_{0\phi}=0,\\
\ee
and the solution of above equation is $A_{0\phi}=-\cos{\th}$ correspond to spilt monopole solution. We obtain an interesting phenomenon: the uniform magnetic field solution can't exist for the GS equation in the  Kerr-Sen BH. To obtain perturbation solution for the toroidal magnetic field and poloidal electric current, we substitute \eqref{13} into \eqref{9}, resulting in
\be
\label{15}
B^{\phi}_1=-\frac{i_1 r(r+r_2)+(2Mr\omega_1-M^2)\sin^2{\th}}{r(r+r_2)\(r(r+r_2)-2Mr\)\sin^2{\th}}.\\
\ee

In the Kerr-Sen BH event horizon with $a=0$ given by $r_{+} = 2M - r_{2}$, a continuity requirement for the toroidal magnetic field constrains the numerator in Eq.\eqref{15} to be zero. Consequently, the first-order electric current and magnetic field are expressed as
\be
\1\{\begin{split}
\label{16}
&i_1= \frac{\sin ^2{\theta} \left(M^2-2M\omega_1(2M-r_2)\right)}{2M(2M-r_2)}\\
&B^{\phi}_1=-\frac{M^2 (2 r (r_2-2 M)\omega_1+Mr+2 M^2)}{2 r^2 (2 M-r_2) (r_2+r)}.
\end{split}\2.
\ee

By expanding the GS equation in terms of the spin parameter $a$, we derive the $o(a^2)$ perturbation solution
\be
\label{17}
\tilde{L}A_{2\phi}=S(r,\theta),\\
\ee
where the source term $S(r,\theta)$ is
\be
\label{18}
\begin{split}
S(r,\theta)&=\frac{\sin ^2{\theta } \left(4 M^2+2 M r+r^2+r r_2\right) \omega'_1}{2 M r \left(r+r_2\right) \left(2 M-r_2\right)}-\frac{\sin {\theta } \cos {\theta } \left(M-2 \left(2M-r_2\right) \omega_1\right){}^2}{2 M^2 \left(2M-r_2\right){}^2}\\
&+\sin {\theta} \cos{\theta} \left(-\frac{M (2 M+r) \left(4 M^2+r_2 (r-2 M)+r^2\right)}{r^2 \left(r+r_2\right){}^2 \left(2M-r_2\right){}^2}+\frac{2 \omega^2_1}{M^2}+\frac{4 (2 M+r) \omega_1  }{r \left(r+r_2\right) \left(2 M-r_2\right)}\right),
\end{split}
\ee
here $'$ represents the derivative with respect to $\theta$. According to the convergence condition proposed by BZ, the restriction need to be imposed on the source \cite{10.1093/mnras/179.3.433}. 
\be
\label{19}
\int^{\infty}_{2M-r_2}\dif r\int^{\pi}_{0}\dif \theta \frac{|S(r,\theta)|}{r}\longrightarrow \text{convergence}.
\ee

It can be demonstrated that to satisfy the Eq. \eqref{19}, the following constraint must be applied
\be
\label{20}
\frac{\sin {\theta } \cos {\theta } \left(\left(8 M-4 r_2\right)  \omega_1-M\right)}{2 M \left(2M-r_2\right){}^2}+\frac{\sin ^2{\theta }  \omega'_1}{2M(2M-r_2)}=0,
\ee
consequently, we have
\be
\1\{\begin{split}
\label{21}
&\omega_1=\frac{M}{4 (2 M-r_2)}\\
&i_1=\frac{M \sin ^2{\theta }}{8 M-4 r_2}=\omega_1\sin ^2{\theta }\\
&B^{\phi}_1=-\frac{M^3 (4 M+r)}{4 r^2 (2 M-r_2) (r_2+r)}.
\end{split}\2.
\ee
Additionally, the second-order component of $A_{\phi}$ can be obtained through 
\be
\label{22}
\tilde{L}A_{2\phi}=-\frac{2 M^2 \sin {\theta } \cos {\theta }}{r^2 (2 M-r_2) (r_2+r)}-\frac{2 M^2 \sin {\theta } \cos {\theta }}{r^2 (r_2+r)^2}.\\
\ee

We can demonstrate that as the parameter $r_2$ approaches 0, the result returns to the Kerr BH. Although the second-order electromagnetic potential $A_{2\phi}$ itself does not alter certain fundamental physical quantities directly, its derivation is essential for a rigorous and high-precision analysis of the BZ process in Kerr-Sen BH for the following reasons. First, the primary motivation is to achieve a higher-order consistent calculation of the energy extraction rate and the extraction efficiency. In strong-field regimes or for BHs with significant spin, first-order approximations are insufficient to capture the subtle energy flux corrections, and second-order terms are required to provide the precision necessary for potential astrophysical comparisons. Second, obtaining the second-order solution allows us to accurately characterize the distribution of magnetic field lines. Specifically, it enables us to manifest how the dilaton charge influences the magnetospheric geometry and the deformation of the field lines beyond the standard Kerr metric. By including these higher-order effects, we can better understand the coupling between the dilaton field and the electromagnetic structure. Therefore, the second-order electromagnetic potential solution in Eq. \eqref{22} is obtained in Appendix A by employing the Green’s function method. From the structure of the solution, it is evident that the presence of the dilaton charge modifies the surrounding magnetic field configuration, giving rise to an additional contribution compared with the Kerr case.

\subsection{The energy extraction rate, radiative efficiency and ergoregion}
With the first-order analytical expressions for $\Omega$ and $I$, we now examine physically observable quantities, specifically the energy extraction rate and efficiency.

The energy extraction rate is defined by
\be
\label{23}
P_{\text{BZ}}=-2\pi \int^{\pi}_{0}\dif \theta \sqrt{-g}T^{r}_{t},
\ee
where $T^{r}_{t}$ is the radial component of the Poynting vector, and the integral is evaluated at surface $r = \text{const}$. Equation \eqref{23} remains independent of the chosen radial coordinate for integration and the configuration of the magnetic field. By utilizing the expression for the energy-momentum tensor of the electromagnetic field, we can calculate the energy extraction rate
\be
\label{24}
\begin{split}
P_{\text{BZ}}&=2\pi\int^{\pi}_{0}I(A_{\phi})\Omega(A_{\phi})\dif A_{\phi}\\
&\approx\frac{2\pi a^2}{M^4} \int^{\pi}_{0}i_1\omega_1\dif A_{0}=\frac{\pi a^2}{6M^2(2M-r_2)^2}=\frac{2\pi}{3}\Omega^{(1)2}_{H},
\end{split}
\ee
where $\Omega^{(1)}_{H}=\frac{a}{2M(2M-r_2)}$ is one-order angular velocity. It is obvious that this result converges to the Kerr BH when the dilation parameter tends to zero. Furthermore, it is noteworthy that in the context of first-order perturbation, the expression $\Omega = \frac{\Omega_{H}}{2}$ approximately aligns with the behavior exhibited by the Kerr BH. Subsequently, we plotted the ratio of two BH extraction rates. The top-left panel in Fig.2 shows the radio of $P_{\rm BZ}$ from different BHs as a function of $r_2$. We observe that within the theoretical range, as the dilaton parameter $r_2$ increases, the Kerr-Sen BH extract several times more energy compared to Kerr BH. In the top-right panel of Fig.2, the ratio of angular velocitie between the two BHs is presented. We find that the angular velocity is sensitive to the dilaton parameter, specifically showing an upward trend as $r_2$ increases for a given spin $a$.

Additionally, the energy extraction efficiency can be expressed as \cite{10.1093/mnras/179.3.433}
\be
\label{25}
\bar{\epsilon}=\frac{\int I\Omega\dif A_{\phi}}{\int I\Omega_{H}\dif A_{\phi}}\approx0.5.
\ee

It is evident that the energy extraction efficiency of the Kerr-Sen BH remains consistent with Kerr BH. Hence, we can't differentiate between these two type of BHs based on the extraction efficiency. The resolution to this issue may potentially emerge from higher-order approximate solution.

Furthermore, the radiative efficiency is also a crucial observational parameter in astrophysics. In the context of BH accretion, it serves as a direct probe of the spacetime geometry near the event horizon and the underlying accretion physics. Higher radiative efficiencies typically indicate that the inner edge of the accretion disk extends closer to the BH, which is strongly influenced by the BH's spin and the accretion mode. In our previous work, we analyzed the radiative efficiency of the Kerr-Sen BH within the framework of the Novikov-Thorne model \cite{Novikov:1973kta,Feng:2023iha}. The Kerr-Sen BH exhibits a markedly higher radiative efficiency compared to its Kerr counterpart. This enhancement implies that, in principle, observational measurements of the radiative output from accretion disks could serve as an effective diagnostic tool for distinguishing between these two classes of BHs.

Moreover, the behavior of $P_{\text{BZ}}$ can easily be understood in terms of the size of the ergoregion, which is the exterior region in which $g_{tt} > 0$, and static observers (time-like and null-like geodesics) are not allowed. The outer and inner boundary of the ergoregion are called the static limit and can be defined by the largest root of $g_{tt} = 0$, namely
\be
\label{26}
\tilde{\Delta}-a^2\sin^2{\th}=0.
\ee

The bottom-right panel in Fig.3 shows the plane $(r\sin{\th}, r\cos{\th})$ of the spacetime. We set $a = 0.5M,0.7M$ and plotted the static limit of the Kerr-Sen BH with $r_2=0$, $0.15M$, and $0.3M$. It can be shown that both the outer and inner boundaries of the ergoregion extend beyond the corresponding event horizons. Furthermore, for the Kerr-Sen BH, the radii of the inner and outer horizons, as well as those of the ergoregion boundaries, are systematically smaller than in the Kerr BH. This configuration ensures the capability of the BH to extract energy from the ergoregion region. The presence of the ergoregion is crucial for the utilization of the BZ mechanism to extract energy. Without this region, even a rotating BH would be unable to extract energy \cite{Komissarov:2004ms,Ruiz:2012te,Toma:2014kva}.
\begin{figure}[H]
\centering
\begin{minipage}{0.5\textwidth}
\centering
\includegraphics[scale=0.8,angle=0]{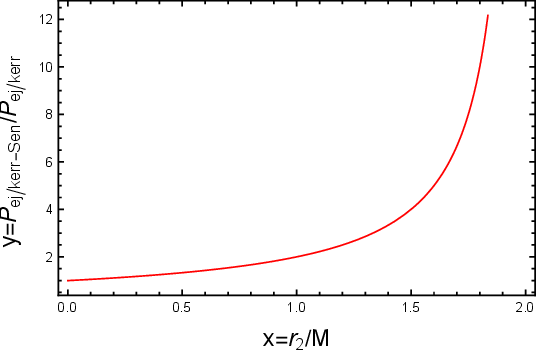}
\end{minipage}%
\begin{minipage}{0.5\textwidth}
\centering
\includegraphics[scale=0.8,angle=0]{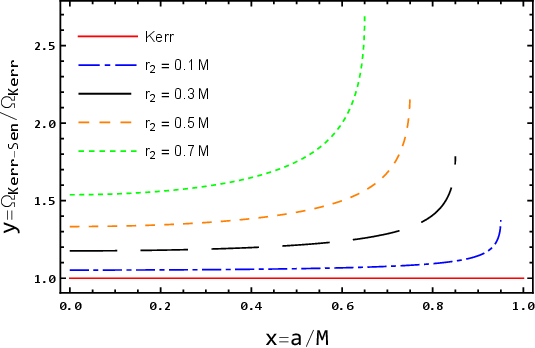}
\end{minipage}
\caption{\label{Fig.2}The left graph illustrates the ratio of energy extraction rate for the two BHs as a function of dilaton parameter $r_2$. The right graph depicts the ratio of angular velocity for a fixed $r_2$ as a function of spin $a$. }
\end{figure}

\begin{figure}[H]
\centering
\begin{minipage}{0.5\textwidth}
\centering
\includegraphics[scale=0.8,angle=0]{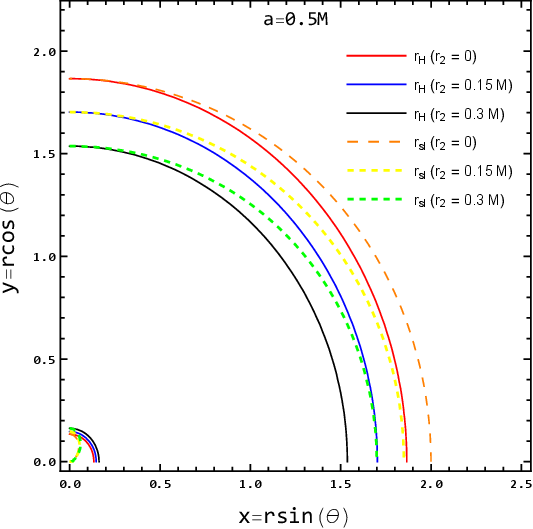}
\end{minipage}%
\begin{minipage}{0.5\textwidth}
\centering
\includegraphics[scale=0.8,angle=0]{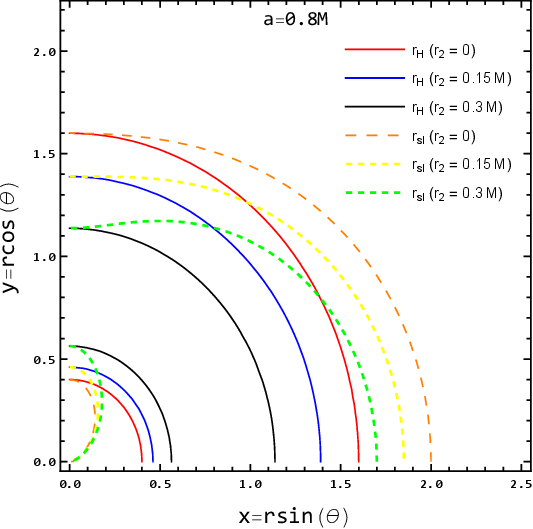}
\end{minipage}
\caption{\label{Fig.3}The bottom panel shows the shape of the ergoregion of the Kerr-Sen BH for $a = 0.5M$ and $0.7M$ with different values of $r_2$ in the $(r \sin\theta,\, r \cos\theta)$ plane.
}
\end{figure}

\section{Comparison with observations}
As discussed in preceding section, the radiative efficiency and the jet power exhibit sensitivity to the metric. Consequently, if observational constraints on these quantities are available for certain Kerr-Sen BH, it provides an avenue to glean insights into the observationally preferred value of the dilaton parameter $r_2$. We will examine six binary BH systems, where both the transient jet power and the BH spin are estimated using the continuum-fitting method. This includes the four objects detailed in Ref. \cite{Narayan:2011eb}: GRS1915+105, GROJ1655-40, XTEJ1550-564, and A0620-00. The fifth source, H1743-322, is discussed in Ref. \cite{steiner2012jet}. The sixth source, GRS 1124-683, has its jet power estimate documented in Ref. \cite{steiner2012jet}, while the spin measurement has been recently conducted using the continuum-fitting method \cite{Chen:2015mvc}. Moreover, the jets from these six systems are known to be mildly relativistic. For these BHs, the bulk Lorentz factor $\Gamma$ is essential for correcting the relativistic beaming when estimating intrinsic jet power from radio luminosity. Observational and statistical studies of these sources suggest that their Lorentz factors typically lie within $2 \lesssim \Gamma \lesssim 5$ \cite{Fender:2004gg,2006csxs.book..381F}. We therefore adopt $\Gamma = 2$ as a conservative lower limit and $\Gamma = 5$ as a representative upper bound, ensuring that our comparison between the Kerr and Kerr–Sen metrics remains physically consistent across the observationally supported range.

To calculate the jet power, we adhere to the methodology outlined in Ref. \cite{Narayan:2011eb}. It is determined by considering the monochromatic flux density at 5 GHz, which is then adjusted for the source's distance, de-boosted based on assumed values for the bulk Lorentz factor ($\Gamma=2$ and $\Gamma=5$), and subsequently normalized by the BH mass to eliminate any dependency \cite{steiner2012jet}. The emitted flux density value, denoted as $S_{\nu,0}$ are shown in Refs. \cite{steiner2012jet,Middleton:2014cha} and presented in Tab.1. For convenience, the corresponding jet powers are provided in Tab.2.

We highlighted that the observed radiative efficiency and jet power can serve as indicators for discerning the preferred value of the dilaton parameter from observational data. To deepen our understanding, we emphasize that a comparison between the theoretical radiative efficiency and jet power with the corresponding observations of six binary BHs (illustrated in Tab.1 and Tab.2). They reveal several consistent patterns: (1) The observed jet power aligns with nearly the entire permissible range of dilaton parameter $r_2$. (2) A higher value $r_2$ necessitates a lower $a$ to account for the observed $P_{\text{jet}}$ and $\epsilon$. (3) The observational constraint on $r_2$ emerges when attempting to replicate the observed $\epsilon$. (4) In most instances, when $r_2=0$ (representing the GR scenario), the observationally permissible range of $a$, obtained from $P_{\text{jet}}$ and $\epsilon$ demonstrates an overlap \cite{PhysRevD.103.044046}.

The characteristics prompt us to assess the chi-square as a variable dependent on $r_2$, achieved through a comparison of $P_{\text{jet}}$ and $\epsilon$ with their respective observational counterparts. This corresponds to the $\chi^2$ expressed as
\be
\label{27}
\chi^2(i,k)=\sum_{j}\frac{(P_{jet,j}-P_{BZ,j}(i,k))^2}{\sigma^2_{P,j}}+\sum_{j}\frac{(\epsilon_{obs,j}-\epsilon_{j}(i,k))^2}{\sigma^2_{\epsilon,j}}
\ee
where $i=\frac{r_2}{M}$ and $k=\frac{a}{M}$. $P_{\text{jet},j}$ is the observational value, $P_{\text{BZ},j}(i,k)$ is the theoretical prediction. $\sigma_{P,j}$ and $\sigma_{\eta,j}$ represent the uncertainties associated with the determination of jet power and radiative efficiency, respectively, where $j=1,...6$ serves as the source index.

For each $r_2$, we systematically vary $a$ within the permissible range: $-(1-\frac{r_2}{2M})\leqslant\frac{a}{M}\leqslant1-\frac{r_2}{2M}$ (ensuring the existence of the event horizon), and calculate $\chi^2$ as delineated in \eqref{27}. The spin parameter yielding the minimum $\chi^2$ for the selected $r_2$. In our analysis, the uncertainty was determined by averaging the upper and lower bounds of the radiative efficiency error. The error range of $P_{\text{jet}}$ was then roughly estimated using formula $P_{\text{jet}}=\(\frac{\nu}{5\text{GHz}}\)\(\frac{S^{\text{max}}_{\nu,0}}{\text{JY}} \)\(\frac{D}{\text{kpc}} \)^2\(\frac{M}{M_{\bigodot}}\)^{-1}$ and adopted as the uncertainty in the transient jet power. In the calculations, the error in the peak 5 GHz radio flux density $\(S_{\nu,0}\)_{\text{max,5GHz}}$ was set to zero. Iterating this process for all $r_2$ values in the range $0\leqslant\frac{r_2}{M}\leqslant2$, we obtain the variation of $\chi^2$ in the Fig.4.
\begin{table}[H]
    \centering
     \caption{\label{Tab.1} Observational values of parameters for six binary BHs are depicted in the table. $D$ represents the distance from the source to the observer, $i^{\circ}$ denotes the inclination angle, and $\epsilon$ signifies the radiative efficiency.}
    \setlength{\tabcolsep}{0.7mm}{
    \begin{tabular}{|llccccc|}
        \hline
        \hline
        BH Binary &  $\frac{a}{M}$  &  $M(M_{\odot})$  &  $D(\text{kpc})$  &  $i^{\circ}$  &  $\epsilon$     &   $(S_{\nu,0})_{\text{max,5GHz}}(\text{Jy})$  \\
        \hline
        A0620-00  & $0.12\pm0.19$                         & $6.61\pm0.25$               & $1.06\pm0.12$                             &  $51.0\pm0.9 $         &$0.061^{+0.009}_{-0.007}$             &  0.203                                                                         \\
        H1743-322  & $0.2\pm0.3$                           & 8                                   &   $8.5\pm0.8$                               &  $75.0\pm3.0$          & $0.065^{+0.017}_{-0.011}$            &  0.0346                                                                      \\
        XTE J1550-564  & $0.34\pm0.24 $               & $9.10\pm0.61$              &  $4.38\pm0.5$                               &  $74.7\pm3.8$           & $0.072^{+0.017}_{-0.011}$          &  0.265                                                                          \\
       GRS 1124-683  & $0.63^{+0.16}_{-0.19}$   & $11.0^{+2.1}_{-1.4}$    & $4.95^{+0.69}_{-0.65}$               & $50.5\pm6.5$           &$0.095^{+0.025}_{-0.017}$            &  0.45                                                                        \\
        GRO J1655-40  & $0.7\pm0.1$                    & $6.30\pm0.27$                & $3.2\pm0.5$                                 & $70.2\pm1.9$           &$0.104^{+0.018}_{-0.013}$            &  2.42                                                                     \\
       GRS 1915+105  & $0.975(>0.95)$                & $12.4^{+1.7}_{-1.9}$    & $8.6^{+2.0}_{-1.6}$                     & $60.0\pm5.0$           &   $0.224(>0.190)$                           &  0.912                                                                           \\
       \hline
    \end{tabular}}
   \end{table}
   \begin{table}[H]
    \centering
     \caption{\label{Tab.2} Jet power proxy values in units of $\text{kpc}^2$ GHz Jy $M^{-1}_{\odot}$}
     \setlength{\tabcolsep}{12mm}{
  \begin{tabular}{|ccc|}
     \hline
        \hline
 BH Binary &  $\Gamma=2|P_{\text{jet}}$    &    $\Gamma=5|P_{\text{jet}}$  \\
\hline
A0620-00            &0.13     &1.6   \\
H1743-322          &7.0       &140  \\
XTE J1550-564   &11        &180  \\
GRS 1124-683    &3.9       &390  \\
GRO J1655-40    &70        &1600 \\
GRS 1915+105   &42        &660   \\
\hline
\end{tabular}}
\end{table}

The Fig.4 illustrates the variation of the logarithm of the $\chi^2$ computed through the aforementioned procedure with the dilaton parameter $r_2$. The red and blue dashed lines represent situations where $P_{\text{BZ}}$ is being compared with the observed $P_{\text{jet}}$ corresponding to $\Gamma=2$ and $\Gamma=5$, respectively. For bulk Lorentz factors $\Gamma = 2$ and $\Gamma = 5$, the best-fit dilaton parameter converges to zero. This outcome suggests that observations of astrophysical BHs are more consistent with the Kerr solution, as the optimal dilaton parameter for Kerr-Sen BH is found to vanish. Consequently, these systems appear to align more closely with the predictions of GR.

\begin{figure}[H]
\centering
\begin{minipage}{0.8\textwidth}
\centering
\includegraphics[scale=1,angle=0]{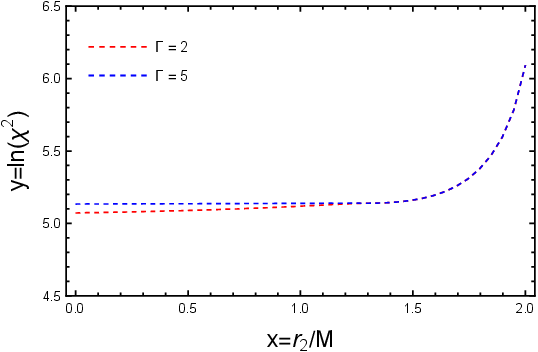}
\end{minipage}%
\caption{\label{Fig.4} The graph illustrates the variation of $\chi^2$ with the dilaton parameter $r_2$ for six binary BHs. The red dashed line represents the situation when $P_{\text{jet}}$ corresponding to $\Gamma=2$ is used to calculate the $\chi^2$, while the blue dashed line is associated with the scenario when $P_{\text{jet}}$ corresponding to $\Gamma=5$ is considered for evaluating the $\chi^2$. }
\end{figure}

\section{Conclusion and discussion}

We have investigated modifications to the BZ mechanism for BHs in alternative theories of gravity. As a concrete example, we explored its implications in the Kerr–Sen BH; a similar analysis could be extended to other non‑Kerr BHs. We present a comprehensive framework for deriving perturbative solutions of the Kerr–Sen BH within the BZ mechanism. Under the assumptions of a stationary, axisymmetric, force‑free magnetosphere, the energy‑momentum conservation equations separate into two constraint functions—the field angular velocity $\Omega(A_{\phi})$ and the poloidal current $I(A_{\phi})$—together with GS equation. The GS equation is a second‑order differential equation requiring two boundary conditions. We impose the horizon‑regularity condition and the convergence condition as the necessary boundaries. With these, all perturbative physical quantities—the magnetic field angular velocity $\Omega$, the current $I$, and the electromagnetic potential $A_{\phi}$—are determined.

Furthermore, we investigated the energy extraction rate and energy extraction efficiency for both the Kerr-Sen and Kerr BHs. Our investigation reveals that as the dilaton parameter $r_2$ increases within the theoretically allowed range, the Kerr-Sen BH exhibits higher energy extraction rate compared to the Kerr BH. However, the energy extraction efficiency of the Kerr-Sen BH remains consistent with Kerr BH. Additionally, the event horizon of the Kerr-Sen BH are smaller than the  boundary of ergosphere, which ensure the BZ mechanism operates reasonably. Finally, we consider six binary BHs by establishing chi-square distribution simulations between the theoretical predictions and observational values of radiative efficiency and energy extraction rate. We observe that the Kerr BH is more consistent with astronomical observation with the bulk lorentz factor $\Gamma=2$ and $\Gamma=5$. This result can be corroborated through the inclusion of a more extensive observational data or by incorporating additional observations in the electromagnetic spectrum, such as quasi-periodic oscillations or BH shadows, which will be detailed in a future work.

\begin{acknowledgments}
This study was supported by the National Natural Science Foundation of China (Grant No. 12333008 ) and the Basic Research Program of Shanxi Province (Grant No. 202503021211204).
\end{acknowledgments}

\appendix
\section{Green's function for solving electromagnetic potential}
The Green's function is pivotal in addressing nonlinear equation. In this appendix, we will follow the approach of  Blandford Znajek and apply it to investigate Kerr-Sen BH. To solve Eq.\eqref{22}, we first consider the Green function, which is given by
\be
\label{A1}
\tilde{L}G=\delta{(r-r_0)}\delta{(\th-\th_0)},
\ee
where the $G$ needs to satisfy boundary conditions such that $G$ is finite at $r=2M-r_0$ and tends to zero as $r\rightarrow \infty$. Its general form solution can be provided as
\be
\label{A2}
G(r,\th;r_0,\th_0)=\sum^{\infty}_{l=0}\frac{l+3/2}{(l+1)(l+2)}\sin{\th}\sin{\th_0}P'_{l+1}(\cos{\th})P'_{l+1}(\cos{\th_0})R_{l}(r,r_0),
\ee
where $P_{l}(\cos{\th})$ is Legendre polynomial and $'$ represents differentiation with respect to $\cos{\th}$. The radial function $R_{l}(r,r_0)$ is
\be
\label{A3}
R_{l}(r,r_0)=\left\{
\begin{aligned}
&U_{l}(r)V_{l}(r_0),2M-r_2\leqslant r\leqslant r_0\\
&V_{l}(r)U_{l}(r_0), r\geqslant r_0\geqslant2M-r_0,\\
\end{aligned}
\right.
\ee
with
\be
\1\{\begin{split}
\label{A4}
&U_{l}(r)=(r+r_2)^2(2M)^lP^{(2,0)}_{l}\(1-\frac{r+r_2}{M}\)\\
&V_{l}(r)=U_{l}(r)\int^{\infty}_{r}\frac{(r'+r_2)\dif r'}{(2M-r'-r_2)U^2_{l}(r')}.\\
\end{split}\2.
\ee

Here, $r_0$ represents a reference point in Kerr-Sen spacetime, $U_{l}(r)$ and $V_{l}(r)$ denote the two solutions correspond to different boundary conditions. $P^{(2,0)}_{l}$ represents Jacobi polynomial. From Eq. (22), we obtain the solution to $\tilde{L}A_{2\phi}=S(r,\theta)$ that vanishes as $r\rightarrow\infty$ and remains regular at the horizon.
\be
\label{A5}
\begin{split}
A_{2\phi}(r,\th)&=\int^{\infty}_{2M-r_2}\dif r_0 \int^{\pi}_{0}\dif \th_0 G(r,\th;r_0,\th_0)S(r_0,\th_0)\\
&=\sin^2{\th}\cos{\th}f(r,r_2),
\end{split}
\ee
with
\be
\1\{\begin{split}
\label{A6}
&f(r,r_2)\equiv\int^{\infty}_{2M-r_2}\(\frac{-2M^2}{r^2_0\(r_0+r_2\)^2}-\frac{2M^2}{r^2_0\(r_0+r_2\)\(2M-r_2\)}\)R_{1}(r,r_0)\dif r_0\\
&U_{1}(r)=2(3M-2r-2r_2)(r+r_2)^2\\
&V_{1}(r)=-\frac{(r+r_2)^2(3M-2r-2r_2)}{432M^5}\(27\log{\frac{r+r_2}{r-2M+r_2}}+\frac{64M}{3M-2(r+r_2)}\right.\\
&\left.-\frac{2M\(3M+11(r+r_2)\)}{(r+r_2)^2}\).
\end{split}\2.
\ee

Due to the complexity of the aforementioned integration, we introduce three dimensionless quantities, $x_0\equiv\frac{r_0}{2M}$, $x\equiv\frac{r}{2M}$, and $q\equiv\frac{r_2}{2M}$. Notably, we recognize that $q$ is a small parameter satisfied $0\leqslant q\leqslant1$, allowing for expansion. Consequently, we have
\be
\label{A7}
\begin{split}
A_{2\phi}(r,\th)&=A^{Kerr}_{2\phi}(x)+\frac{q}{72}\(w_{1}(x)+w_2(x)\)+o(q^2),
\end{split}
\ee
with
\be
\1\{\begin{split}
\label{A8}
&A^{Kerr}_{2\phi}(x)\equiv\frac{11}{72}+\frac{1}{6x}+x-2x^2-\frac{\pi^2 x^2}{4}+\frac{\pi^2x^3}{3}+\frac{\log{x}}{12}+\frac{x\log{x}}{2}-2x^2\log{x}\\
&-\frac{3x^2(\log{x})^2}{4}+x^3(\log{x})^2+2x^3Li_{2}(1-x)-\frac{3x^2Li_{2}(1-x)}{2}\\
&w_{1}(x)\equiv151-\frac{3}{x^2}+\frac{30}{x}-144x-36\pi^2x-432x^2-18\pi^2x^2+120\pi^2x^3\\
&+36x(-6+x(20x-3))Li_{2}(1-x)\\
&w_2(x)\equiv\frac{360 x^4 \log ^2(x)}{x-1}+\frac{144 x^4 \log (x)}{x-1}-\frac{414 x^3 \log ^2(x)}{x-1}-144 x^3 \log (x-1)\\
&+432 x^3 \log \left(\frac{x-1}{x}\right)+\frac{2052 x^3 \log (x)}{x-1}-\frac{54 x^2 \log ^2(x)}{x-1}-2916 x^2 \log (x-1)\\
&+2700 x^2 \log \left(\frac{x-1}{x}\right)-\frac{3960 x^2 \log (x)}{x-1}+\frac{108 x \log ^2(x)}{x-1}+1512 x \log (x-1)\\
&-1512 x \log \left(\frac{x-1}{x}\right)+\frac{1794 x \log (x)}{x-1}-\frac{66 \log (x)}{x-1}.
\end{split}\2.
\ee
where the function $Li_{2}(x)$ is the dilogarithm function, also known as the Spence's function. It is defined as
\be
\label{A9}
Li_{2}(x)=-\int^{x}_{0}\frac{\ln(1-t)}{t}\dif t.
\ee

As $q$ approaches 0, this analytical solution automatically reverts to the second-order correction of the Kerr BH. The influence of the dilation parameter only manifests in the correction term.

\bibliographystyle{unsrt}
\bibliography{EMDAFF}

\end{document}